\documentstyle[amssymb,aps,12pt]{revtex}

\begin{document}
\draft
\author{O. B. Zaslavskii}
\address{Department of Mechanics and Mathematics, Kharkov V.N. Karazin's National\\
University, Svoboda\\
Sq.4, Kharkov 61077, Ukraine\\
E-mail: ozaslav@kharkov.ua}
\title{Near-extremal and extremal quantum-corrected two-dimensional charged black
holes}
\maketitle

\begin{abstract}
We consider charged black holes within dilaton gravity with
exponential-linear dependence of action coefficients on dilaton and minimal
coupling to quantum scalar fields. This includes, in particular, CGHS and
RST black holes in the uncharged limit. For non-extremal configuration
quantum correction to the total mass, Hawking temperature, electric
potential and metric are found explicitly and shown to obey the first
generalized law. We also demonstrate that quantum-corrected extremal black
holes in these theories do exist and correspond to the classically forbidden
region of parameters in the sense that the total mass $M_{tot}<Q$ ($Q$ is a
charge). We show that in the limit $T_{H}\rightarrow 0$ (where $T_{H}$ is
the Hawking temperature) the mass and geometry of non-extremal configuration
go smoothly to those of the extremal one, except from the narrow
near-horizon region. In the vicinity of the horizon the quantum-corrected
geometry (however small quantum the coupling parameter $\kappa $ would be)
of a non-extremal configuration tends to not the quantum-corrected extremal
one but to the special branch of solutions with the constant dilaton (2D
analog of the Bertotti-Robinson metric) instead. Meanwhile, if $\kappa =0$
exactly, the near-extremal configuration tends to the extremal one. We also
consider the dilaton theory which corresponds classically to the
spherically-symmetrical reduction from 4D case and show that for the
quantum-corrected extremal black hole $M_{tot}>Q$.
\end{abstract}



\section{Introduction}

Two-dimensional (2D) models of dilaton gravity serve as an excellent tool
which enables us to understand better the crucial points of black hole
physics \cite{callan} in a more simple context than in the four-dimensional
(4D) case (see for recent reviews \cite{od}, \cite{dv}). One of such issues
is the role and character of backreaction caused by quantum fields. This
backreaction reveals itself for both non-extremal black holes (NEBH) and
extremal ones (EBH). In the first case it affects, in particular, the value
of the Hawking temperature.

For the CGHS black hole \cite{callan} the corresponding correction was
calculated in \cite{pl} for the Hartle-Hawking state and, in an independent
approach, in \cite{pos} for the evaporating black hole. Further, the
corresponding results were generalized to take into account the presence of
a point-like shell that is necessary for the black hole to maintain thermal
equilibrium with its Hawking radiation \cite{2dwall} (the influence of a
shell on thermodynamics of 4D quantum-corrected black holes was considered
in \cite{bou}). Quantum corrections to the geometry and thermodynamics of
NEBH were also considered for another model which classically arises in the
''spherically-symmetrical gravity'' (SGG) theory as a result of
spherically-symmetrical reduction from 4D case \cite{sol3}.

Consideration of \cite{sol3} was generalized in \cite{gab} and \cite{bur}
but the issue of relationship between quantum-corrected NEBH and EBH
remained not completely clear in what concerns the limiting transition and
comparison of physical quantities like mass and geometry. It is worth
mentioning that in recent years the very existence of semiclassical EBH in
general relativity became the subject of discussion \cite{lowe}, \cite{paul3}%
. Direct treatment of quantum backreaction caused by massive fields on the
Reissner-Nordstr\"{o}m black hole \cite{jurek} confirmed that the
semiclassical EBH do exist in general relativity, with the extremality
condition slightly modified due to quantum corrections. This conclusion was
also extended to 2D generic dilaton gravity \cite{ecq}.

The aim of the present paper is to find in the explicit form quantum
corrections to the metric and thermodynamic quantities of NEBH and trace in
detail the extremal limit $T_{H}\rightarrow 0$ ($T_{H}$ is the Hawking
temperature) from quantum-corrected NEBH. We will see that in the
near-horizon region the NEBH approaches not EBH but a special branch of
solution with the constant dilaton field which is the 2D analogue of
Bertotti-Robinson spacetime. We find quantum corrections to the mass of EBH
and show that it agrees with the limiting value of NEBH. We also calculate
quantum corrections to the mass of the EBH in the SGG model. It turns out
that $M_{tot}<Q$ for the quantum-corrected CGHS charged black hole but $%
M_{tot}>Q$ for the latter case.

It is worth noting that in typical exactly solvable models always $%
T_{H}=const\neq 0$ \cite{exact}. Therefore, generic EBH (except some special
classes with vanishing potentials \cite{ecq} which have nothing to do with
the models discussed in the present paper) do not fall into the family of
exactly solvable models. In such a situation, one is led to using a
perturbative approach with respect to the quantum-coupling parameter $\kappa 
$. In doing so, one should be careful about the choice of the unperturbed
state since, as is aforementioned, the relationship between the mass and
charge for the quantum-corrected EBH can be such that classical EBH with the
same $M_{tot}$ and $Q$ are forbidden at all. It means that the problem
should be considered in a self-consistent approach from the very beginning,
with the behavior of the metric ensuring the condition $T_{H}=0$.

To avoid complications which are unnessary for the issues under
consideration we discuss only backreaction caused by scalar field minimally
coupled to the dilaton, so it is decribed by the Polyakov action. (The issue
EBH in theories with the non-minimal coupling is the separate subject which
is beyond the scope of the present paper. We only mention that rather
recently the existence of EBH solutions was confirmed with the help of
numeric calculations in \cite{fab}.)

Throughout the paper we use the Schwarzschild gauge instead of that used for
studying quantum corrections in \cite{sol3} for SGG model and in \cite{gab}
for a more general case. In doing so, we do not fix in advance the conformal
frame, so all coefficients in the action (\ref{ac}) are generic with the
only constraint that classically the linear dilaton vacuum is the exact
solution of field equations. To avoid complication connected with a finite
size, in this paper we restrict ourselves to the system in an infinite space.

The paper is organized as follows. In Sec. 2 we list field equations for the
quantum-corrected gravitation-dilaton system and suggest the simplified
approach in which we find the general expression for the correction to $%
T_{H} $ in terms of classical system directly from field equations. In Sec.
3 we apply the previous formulas to the charged version of CGHS black hole
with modification of dilaton potentials which for the uncharged version
ensured exact solvability. The explicit expressions for the metric, mass,
temperature, potential and entropy are obtained in terms of the horizon
value of the dilaton and charge are obtained and the validity of the first
law is verified. In Sec. 4 we carry out independent consideration of
semiclassical extremal black holes which from the very beginning fall into
another topological sector. We find, how the relationship $M=Q$ between the
mass and charge, typical of classical EBH, modifies due to quantum
corrections. In Sec. 5 we discuss what happens to the metric of the
quantum-corrected near-extremal NEBH in the immediate neighborhood of the
horizon and show the qualitative difference here with the pure classical
case. In Sec. 6 we repeat briefly calculations of the mass of EBH for the
quantum-corrected SGG model. Sec. 7 contains summary and discussion of the
results.

\section{Field equations}

Let us consider the system governed by the action 
\begin{equation}
I=I_{gd}+I_{PL}+I_{q}\text{, }
\end{equation}
where gravitation-dilaton part 
\begin{equation}
I_{gd}=\frac{1}{2\pi }\int_{M}d^{2}x\sqrt{-g}[F(\phi )R+V(\phi )(\nabla \phi
)^{2}+U(\phi )]\text{,}  \label{ac}
\end{equation}
the contribution of the electromagnetic field is 
\begin{equation}
I_{q}=-\frac{1}{4\pi }\int_{M}d^{2}x\sqrt{-g}W(\phi )F_{\mu \nu }F^{\mu \nu }%
\text{,}
\end{equation}
$F_{\mu \nu }F^{\mu \nu }=2F_{01}F^{01}\equiv -2E^{2}$and we omit boundary
terms that does not affect the form of field equations.

$I_{PL\text{ }}$is the Polyakov-Liouville action \cite{polyakov}
incorporating effects of Hawking radiation and its backreaction on spacetime
for a multiplet of N minimal scalar fields. It is convenient to write it
down in the form 
\begin{equation}
I_{PL}=-\frac{\kappa }{2\pi }\int_{M}d^{2}x\sqrt{-g}[\frac{(\nabla \psi )^{2}%
}{2}+\psi R]\text{.}  \label{ipl}
\end{equation}
Varying (\ref{ipl}) with respect to $\psi $, it is easy to obtain that the
function $\psi $ obeys the equation 
\begin{equation}
\square \psi =R\text{,}  \label{4}
\end{equation}
where $\Box =\nabla _{\mu }\nabla ^{\mu }$, $\kappa =N/24$ is the quantum
coupling parameter. Varying the action with respect to the metric, we get 
\begin{equation}
T_{\mu \nu }\equiv T_{\mu \nu }^{(gd)}+T_{\mu \nu }^{(q)}+T_{\mu \nu
}^{(PL)}=0\text{,}  \label{6}
\end{equation}
where 
\begin{equation}
T_{\mu \nu }^{(gd)}=\frac{1}{2\pi }\{2(g_{\mu \nu }\square F-\nabla _{\mu
}\nabla _{\nu }F)-Ug_{\mu \nu }+2V\nabla _{\mu }\phi \nabla _{\nu }\phi
-g_{\mu \nu }V(\nabla \phi )^{2}\}\text{,}  \label{7}
\end{equation}
\begin{equation}
T_{\mu \nu }^{(PL)}=-\frac{\kappa }{2\pi }\{\partial _{\mu }\psi \partial
_{\nu }\psi -2\nabla _{\mu }\nabla _{\nu }\psi +g_{\mu \nu }[2\square \psi -%
\frac{1}{2}(\nabla \psi )^{2}]\}\text{,}  \label{8}
\end{equation}
\[
T_{\mu \nu }^{(q)}=-\frac{W}{2\pi }(g_{\mu \nu }E^{2}+2g^{\alpha \beta
}F_{\mu \alpha }F_{\nu \beta })\text{.} 
\]
Variation of the action with respect to $\phi $ gives rise to the equation 
\begin{equation}
RF^{^{\prime }}+U^{\prime }+W^{\prime }E^{2}=2V\Box \phi +V^{\prime }(\nabla
\phi )^{2}\text{,}  \label{9}
\end{equation}
where prime denotes derivative with respect to $\phi $.

Let us write down the metric in the Schwarzschild gauge: 
\begin{equation}
ds^{2}=-fdt^{2}+f^{-1}dx^{2}\text{.}  \label{gauge}
\end{equation}
It is worth noting that field equations of the semiclassical system under
discussion look very much like those for a pure classical one, but with
coefficients, shifted according to (\ref{shift})

\begin{equation}
\tilde{F}=F-\kappa \psi \text{, }\tilde{V}=V-\kappa \frac{\psi ^{\prime 2}}{2%
}\text{.}  \label{shift}
\end{equation}

In this paper we will be dealing with static solutions only. This is
possible due to the fact that the structure of field equations exhibits the
existence of the Killing vector $\xi _{\alpha }=e_{\alpha }^{\beta }\mu
_{;\beta }$, where $\mu ^{\prime }=\tilde{F}^{\prime }\exp (-\int d\phi 
\frac{\tilde{V}}{\tilde{F}^{\prime }})$, which is assumed to be time-like
for the static case. The latter means that the metric, the fields $\psi $
and $\phi $ depend on the coordinate $x$ only. Alternatively, one can use
instead of $x$ the dilaton field $\phi $ itself.

Let us consider a configuration with the fixed charge $Q$. It follows from
Maxwell equations $(WF^{\mu \nu })_{;\nu }=0$ that 
\begin{equation}
F_{01}=-QW^{-1}\sqrt{-g}  \label{f01}
\end{equation}

Then field equations take the following explicit form: 
\begin{equation}
2f\frac{d^{2}\tilde{F}}{dx^{2}}+\frac{df}{dx}\frac{d\tilde{F}}{dx}-U_{eff}-%
\tilde{V}f\left( \frac{d\phi }{dx}\right) ^{2}=0\text{,}  \label{sc00}
\end{equation}

\begin{equation}
\frac{df}{dx}\frac{d\tilde{F}}{dx}-U_{eff}+\tilde{V}f\left( \frac{d\phi }{dx}%
\right) ^{2}=0\text{,}  \label{sc11}
\end{equation}
\begin{equation}
U_{eff}=U-Q^{2}W^{-1}\text{.}  \label{ueff}
\end{equation}
It is also convenient to take the difference of (\ref{sc00}), (\ref{sc11})
to get 
\begin{equation}
\frac{d^{2}\tilde{F}}{dx^{2}}=\tilde{V}\left( \frac{d\phi }{dx}\right) ^{2}
\label{dif}
\end{equation}
or 
\begin{equation}
\left( \frac{d^{2}F}{d\phi ^{2}}-V\right) \left( \frac{d\phi }{dx}\right)
^{2}+\kappa [\frac{1}{2}\left( \frac{d\psi }{dx}\right) ^{2}-\frac{d^{2}\psi 
}{dx^{2}}]+\frac{dF}{d\phi }\frac{d^{2}\phi }{dx^{2}}=0\text{.}  \label{1}
\end{equation}
In the present paper, we will consider the models for which $\frac{d^{2}F}{%
d\phi ^{2}}-V=-\kappa \frac{c}{4}$, where $c$ is a constant (the coefficient 
$\frac{1}{4}$ is singled out for convenience ) and the linear dilaton vacuum 
$\phi =-z\equiv -\lambda x$ is the exact classical solution. This condition
is not very restrictive and includes, for example, the most part of
string-inspired models (see below).

Let us denote by ''+'' sign the quantities calculated on the horizon. Eq. (%
\ref{4}) can be integrated for static metrics to give 
\begin{equation}
\frac{d\psi }{dx}=f^{-1}[\left( \frac{df}{dx}\right) _{+}-\frac{df}{dx}]%
\text{,}  \label{psi}
\end{equation}
where the constant integration is chosen to ensure the regularity of $\psi $
near the horizon of non-extreme black holes, where $f\sim x-x_{+}$.

Then in the first approximation, with the main correction taken into
account, 
\begin{equation}
\phi =-z+\kappa \rho \text{,}  \label{zro}
\end{equation}
\begin{equation}
\frac{dF}{d\phi }\frac{d^{2}\rho }{dx^{2}}+[\frac{1}{2}\left( \frac{d\psi }{%
dx}\right) ^{2}-\frac{d^{2}\psi }{dx^{2}}]=0\text{.}
\end{equation}
With the same accuracy, 
\begin{equation}
\frac{dF}{d\phi }\frac{d^{2}\rho }{d\phi ^{2}}-s=\frac{c}{4}\text{.}
\end{equation}
\begin{equation}
\frac{d\rho }{d\phi }=\int_{-\infty }^{\phi }d\phi \left( \frac{dF}{d\phi }%
\right) ^{-1}(s+\frac{c}{4})\text{.}  \label{ro}
\end{equation}
\begin{equation}
s\equiv -f^{-1}\frac{d^{2}f}{d\phi ^{2}}+\frac{1}{2f^{2}}[\left( \frac{df}{%
d\phi }\right) ^{2}-\left( \frac{df}{d\phi }\right) _{+}^{2}]\text{.}
\label{s}
\end{equation}
At $x\rightarrow \infty $, $\phi \rightarrow -\infty $ we have a linear
dilaton vacuum.

Another field equation is obtained as the sum of eqs. (\ref{sc00}), (\ref
{sc11}): 
\begin{equation}
f\frac{d^{2}F}{dx^{2}}+\frac{df}{dx}\frac{dF}{dx}=U_{eff}-\kappa f^{\prime
\prime }\text{.}  \label{3}
\end{equation}
Rescaling the potential according to $U_{eff}=\lambda ^{2}u$, we have 
\begin{equation}
u=\left( \frac{df}{d\phi }\frac{d\tilde{F}}{d\phi }+f\frac{d^{2}\tilde{F}}{%
d\phi ^{2}}\right) \left( \frac{d\phi }{dz}\right) ^{2}+\frac{d\tilde{F}}{%
d\phi }\frac{d^{2}\phi }{dz^{2}}=\frac{d}{dz}\left( f\frac{d\tilde{F}}{dz}%
\right) \text{.}  \label{uf}
\end{equation}

At the horizon $f=0$ and we obtain directly from (\ref{3}) 
\begin{equation}
T_{H}=\frac{1}{4\pi }\left( \frac{df}{dx}\right) _{+}=\frac{1}{4\pi }\left| 
\frac{U_{eff}-\kappa f^{\prime \prime }}{\frac{dF}{d\phi }\frac{d\phi }{dx}}%
\right| _{+}\text{.}
\end{equation}

Retaining the terms of the first order in $\kappa $, we have $\frac{d\phi }{%
dz}=-(1+\kappa \frac{d\rho }{d\phi })+...$ and 
\begin{equation}
T_{H}=\frac{\lambda }{4\pi }\left( \frac{\partial F}{\partial \phi }\right)
_{+}^{-1}[u_{+}-\kappa (\frac{d^{2}f^{(0)}}{d\phi ^{2}}+u_{+}(\frac{d\rho }{%
d\phi })_{+}]\text{,}  \label{thgen}
\end{equation}
where $f^{(0)}$ is the solution of classical field equations and $(\frac{%
\partial \rho }{\partial \phi })_{+}$ is given buy the integral (\ref{ro}).

As, by assumption, classically 
\begin{equation}
\frac{d^{2}F}{d\phi ^{2}}-V=0\text{,}  \label{ffv}
\end{equation}
it follows from field equations with $\kappa =0$ that $\phi =-z$ and 
\begin{equation}
f^{(0)}=\left( \frac{dF}{d\phi }\right) ^{-1}\int_{\phi _{+}}^{\phi }d\phi u%
\text{.}  \label{fcl}
\end{equation}

By substitution to eq. (\ref{thgen}) we obtain 
\[
T_{H}=T_{H}^{(0)}(1-\kappa D)\text{, }T_{H}^{(0)}=\frac{\lambda }{4\pi }%
\left( \frac{dF}{d\phi }\right) _{+}^{-1}u_{+}\text{, }D=\frac{u_{+}^{\prime
}}{u_{+}F_{+}^{\prime }}-2\frac{F_{+}^{\prime \prime }}{F_{+}^{\prime 2}}+(%
\frac{d\rho }{d\phi })_{+}\text{.} 
\]

\section{Quantum corrections to geometry and thermodynamics}

From now on we will consider the class of models with 
\begin{equation}
F=\exp (-\phi )+\kappa \frac{b}{2}\phi \text{, }V=\exp (-\phi )+\frac{\kappa 
}{4}c\text{, }W=\exp (-\phi )\text{, }u=\exp (-\phi )-q^{2}\exp (\phi )\text{%
, }  \label{FV}
\end{equation}
where we rescaled the charge according to $Q=q\lambda $. 
\begin{equation}
\tilde{F}=\exp (-\phi )+\kappa \frac{b+2}{2}\phi \text{.}  \label{FF}
\end{equation}

In the uncharged case $q=0$ the system under consideration represents
deformation of the CGHS model to which it reduces if $b=c=0$. If $p\equiv b+%
\frac{c}{2}+1=0$, we obtain the CN model \cite{cruz} that contains, as the
particular cases, the RST model ($c=0$, $b=-1$) \cite{rst} and BPP ($c=2$, $%
b=-2$) \cite{bose}. In turn, the CN model is the particular class of a more
wide family of exactly solvable models \cite{exact}.

The classical limit $\kappa =0$ of the model (\ref{FV}) was analyzed in \cite
{frolov}, \cite{gibper}. In this limit

\begin{equation}
f^{(0)}=1-2me^{\phi }+q^{2}e^{2\phi }=1-\frac{2m}{r}+\frac{q^{2}}{r^{2}}=(1-%
\frac{r_{+}}{r})(1-\frac{r_{-}}{r})\text{, }r\equiv \exp (-\phi )\text{.}
\label{fmq}
\end{equation}
The equation $f(r)=0$ has two roots, if $m\geq q$, 
\begin{equation}
r_{\pm }=\exp (-\phi _{\pm })=m\pm \sqrt{m^{2}-q^{2}}\text{, }m=\frac{\exp
(-\phi _{+})+\exp (-\phi _{-})}{2}\text{, }q^{2}=\exp [-(\phi _{+}+\phi
_{-})]\text{,}
\end{equation}
the event horizon corresponding to $r_{+}$. Hereafter, we denote by the
index ''$+$'' the quantities calculated at $r_{+}$. Then the Hawking
temperature $T_{H}=\frac{1}{4\pi }\left( \frac{df}{dx}\right) _{+}$ is equal
to 
\begin{equation}
T_{H}^{(0)}=\frac{\lambda }{2\pi }\frac{\sqrt{m^{2}-q^{2}}}{m+\sqrt{%
m^{2}-q^{2}}}\text{.}
\end{equation}

Now it is the issue of quantum corrections that we turn to. Calculating the
quantity $s$ in (\ref{s}) with respect to the unperturbed metric, we obtain 
\begin{equation}
s=\frac{1}{2}[a_{0}+\frac{a_{1}r_{-}}{r-r_{-}}+\frac{a_{2}r_{-}^{2}}{\left(
r-r_{-}\right) ^{2}}]\text{.}
\end{equation}
\begin{equation}
a_{0}=-(1-x)^{2}\text{, }a_{1}=2(x+1)(2-x)\text{, }a_{2}=1-x^{2}\text{, }
\end{equation}
\begin{equation}
\frac{d\rho }{d\phi }=\frac{1}{2r}[\left( 1-x\right) ^{2}-\frac{c}{2}%
+(1+x)\mu ]\text{,}
\end{equation}
\begin{equation}
\mu =\frac{2\ln (1-y)}{y}+\frac{2+y(x-3)}{1-y}\text{,}
\end{equation}
where 
\[
x=\frac{r_{-}}{r_{+}}\text{, }y=\frac{r_{-}}{r}\text{.} 
\]

Finding the solutions of eq. (\ref{uf}) perturbatively, we obtain

\begin{equation}
f=f^{(0)}+\kappa f^{(1)}\text{,}  \label{f}
\end{equation}

\begin{equation}
f^{(0)}=(1-y)(1-u)\text{, }u\equiv \frac{r_{+}}{r}\text{.}  \label{f0}
\end{equation}
\begin{equation}
f_{1}=\frac{\chi }{r}\text{, }\chi =\chi _{0}+\chi _{1}\text{, }  \label{f1}
\end{equation}
\begin{equation}
\chi _{0}=\frac{(\phi -\phi _{+})}{2}[(1-x)^{2}-\frac{c}{2}]\text{.}
\label{hi}
\end{equation}

\begin{equation}
\chi _{1}=f^{(0)}\frac{p}{2}+\frac{(1-u)}{2}[3y-3-2(x+x^{2})-\frac{c}{4}%
x(1+u)]-2(1-u)(1+x)(1-y)\frac{\ln (1-y)}{y}+\frac{G}{2}\text{.}
\end{equation}
\begin{equation}
G=\frac{(1+x)}{x}(1-x)^{2}\ln [\frac{1-x}{1-y}]\text{.}
\end{equation}

In the afore-mentioned formulas the quantity the horizon ''radius'' $r_{+}$
is supposed to be fixed. For the quantum-corrected metric the geometry near $%
r_{-}$ becomes in general singular (so $r_{-}$ looses its direct meaning) in
agreement with general properties of Cauchy horizons. Then, $r_{-}$ should
be understood as the formal quantity $r_{-}=\frac{q^{2}}{r_{+}}$.

\subsection{Hawking temperature}

Knowing the functions $f(\phi )$ (\ref{f})-(\ref{hi}) and $z(\phi )$ (\ref
{zro}), (\ref{ro}) one can calculate in the same approximation the Hawking
temperature which agrees with (\ref{thgen}) for the model under discussion:

\begin{equation}
T_{H}=T_{0}[1-x+\kappa \eta \exp (\phi _{+})]\text{, }T_{0}\equiv \frac{%
\lambda }{4\pi }\text{.}  \label{tcor}
\end{equation}
\begin{equation}
\eta =-[\frac{x^{2}+x(1+p)+2-p}{2}-\frac{(1+x)(1-x)}{x}\ln (1-x)]\text{,}
\end{equation}
\[
p=b+\frac{c}{2}+1\text{.} 
\]

In the uncharged case $x=0$, $\eta =\frac{1}{2}$ and we return to the result
of \cite{2dwall}. If $\kappa \ll 1-x$, the quantum correction is small as
compared to the classical Hawking temperature. If $\kappa \sim 1-x\ll 1$,
the term with a logarithm can be neglected and the condition of the
extremality changes to 
\begin{equation}
1-x=2\kappa \exp (\phi _{+})\text{.}
\end{equation}
It agrees with the general condition 
\begin{equation}
U(\phi _{+})F^{\prime }(\phi _{+})=\kappa U^{\prime }(\phi _{+})
\label{cond}
\end{equation}
which should hold on the event horizon of quantum-corrected extremal black
holes (see \cite{ecq}, Sec. 8.1).

\subsection{Potential}

If we write $F_{01}=-\frac{\partial \varphi }{\partial \xi }$, then it
follows from (\ref{f01}) that the potential

\begin{equation}
\varphi =-q\int dz\frac{1}{r}=q\int_{-\infty }^{\phi }d\phi (1-\kappa \rho
^{\prime })\exp (\phi )=\frac{q}{r}+\varphi _{1}^{(1)}\text{, }\varphi
_{1}^{(1)}=\kappa q\varphi _{1}
\end{equation}
and after direct calculations we obtain 
\begin{equation}
\varphi _{1}=\frac{a}{2r_{-}^{2}}\text{,}
\end{equation}
\begin{equation}
a(x)=x+x^{3}+\frac{c}{4}x^{2}+(1-x^{2})\ln (1-x).
\end{equation}

In the limit $q\rightarrow 0$, $r_{-}\rightarrow 0$, $x\rightarrow 0$ $%
\varphi _{1}$ does not contain $q$ and we see that $\varphi _{1}^{(1)}\sim
q\rightarrow 0$, as it should be.

\subsection{Energy}

The energy of the classical gravitational-dilaton system with the
string-inspired Lagrangian was obtained in the closed form in \cite{frolov}, 
\cite{gibper}. Its generalization to the generic quantum-corrected case is
direct. The energy (see, for example, \cite{gab}, \cite{found}) is equal to

\begin{equation}
E=-\frac{1}{\pi }\left( \frac{d\tilde{F}}{dl}\right) _{B}=-\frac{1}{\pi }%
\left( \frac{d\tilde{F}}{dx}\sqrt{f}\right) _{B}=\frac{\lambda }{\pi }%
\varepsilon \text{,}  \label{e}
\end{equation}
\begin{equation}
\varepsilon =\varepsilon ^{(0)}+\kappa \varepsilon ^{(1)}\text{.}  \label{e1}
\end{equation}
where $f$ and $\tilde{F}$ can be taken from eqs. (\ref{FF}), (\ref{f}) - (%
\ref{hi}), $dl$ is the proper distance. We are interested in the ADM mass $%
M_{tot}=\lim_{r\rightarrow \infty }(E-E_{0})$, where $E_{0}$ is the energy
of some background configuration.

Calculating (\ref{e}), (\ref{e1}) and collecting all terms of the order $%
\kappa $ together, we obtain at $r\rightarrow \infty $ 
\begin{equation}
\varepsilon ^{(0)}=-r+\frac{r_{+}+r_{-}}{2}\text{.}
\end{equation}
\begin{equation}
\varepsilon _{1}=-\frac{x}{2}+\frac{1}{4}(1-p)+\frac{cx}{16}-\frac{1}{4}%
\frac{(1+x)}{x}(1-x)^{2}\ln (1-x)+\frac{(\phi _{+}-\phi )}{4}(1-x)^{2}+\frac{%
c(\phi -\phi _{+})}{8}
\end{equation}

\subsection{Background}

We choose, as the reference configuration, the ''quasiflat'' metric which
would correspond in the limit $\kappa =0$ to the classical solution $f=1+%
\frac{q^{2}}{r^{2}}$, $\phi =-z+const$ and represents at infinity the vacuum
state with vanishing stresses. Formally, it means that we should replace in
the expression (\ref{psi}) the quantity $\left( \frac{df}{dx}\right) _{+}$
by zero. If $q=0$, one can observe that the linear dilaton vacuum is the
exact solution of one-loop field equations with $b=c=0$ and all possible
corrections come only due to non-zero $b$ and $c$ (''quasiflat''
background). Repeating calculations with $q\neq 0$, we obtain from field
equations (\ref{sc00}), (\ref{sc11}) that at $r\rightarrow \infty $%
\[
f=1+\frac{\kappa }{2r}(b+\frac{c}{2}-\frac{c}{2}\phi )+O(\frac{1}{r^{3}})%
\text{,}
\]
the corresponding contribution to the energy

\begin{equation}
\varepsilon ^{quasiflat}=-\frac{dF}{dz}\sqrt{f}=-r+\kappa \varepsilon _{1}
\end{equation}
\begin{equation}
\varepsilon _{1}^{quasiflat}=-\frac{1}{2}(\frac{p-1}{2}-\frac{c\phi }{4})-%
\frac{(1-p)}{2}
\end{equation}
\begin{equation}
\varepsilon _{1}^{flat}=-\frac{1}{4}(1-p)+\frac{c\phi }{8}
\end{equation}
\begin{equation}
\varepsilon _{0}=\frac{r_{+}+r_{-}}{2}\text{.}
\end{equation}
\begin{equation}
\varepsilon _{1}-\varepsilon _{1}^{flat}=\varepsilon ^{th}-\frac{x}{2}-\frac{%
cx}{16}-\frac{1}{4}\frac{(1+x)}{x}(1-x)^{2}\ln (1-x)-\frac{c\phi _{+}}{8}+%
\frac{1-p}{2}\text{, }
\end{equation}
\begin{equation}
\varepsilon ^{th}=\frac{(\phi _{+}-\phi )}{4}(1-x)^{2}\text{,}
\end{equation}
\begin{equation}
M_{tot}=M_{th}+M_{BH}\text{,}  \label{mtot}
\end{equation}
\begin{equation}
M_{th}=\kappa \lambda \frac{(\phi _{+}-\phi )}{4\pi }(1-x)^{2}=\frac{\pi }{6}%
T_{H}^{2}L\text{, }\lambda L=\phi _{+}-\phi \text{, }T_{H}=T_{0}(1-x)\text{, 
}T_{0}=\frac{\lambda }{4\pi }
\end{equation}
\begin{equation}
M_{BH}=M_{BH}^{(0)}+2\kappa T_{0}M_{1}\text{, }%
M_{BH}^{(0)}=2T_{0}(r_{+}+r_{-})\text{, }M_{1}=-x+\frac{cx}{8}-\frac{1}{2}%
\frac{(1+x)}{x}(1-x)^{2}\ln (1-x)-\frac{c\phi _{+}}{4}
\end{equation}
\begin{equation}
M_{tot}=M_{BH}+T_{0}\kappa (\phi _{+}-\phi )(1-x)^{2}\text{, }M_{BH}=4T_{0}[%
\frac{r_{+}+r_{-}}{2}+\kappa (m_{1}-\frac{c\phi _{+}}{8})]  \label{tot}
\end{equation}
\begin{equation}
m_{1}=-\frac{x}{2}+c\frac{x}{16}-\frac{1}{4}\frac{(1+x)}{x}(1-x)^{2}\ln (1-x)
\end{equation}

In the limit $x\rightarrow 0$ $m_{1}=\frac{1}{4}$ in accordance with \cite
{2dwall}.

It follows from (\ref{mtot})-(\ref{tot}) that, independently of values of $b$
and $c$ the total mass splits to separate contributions $M_{BH}$ and $M_{th}$
which come from a black hole itself and thermal gas between a horizon having
a temperature $T_{H}$.

\subsection{Corrections to entropy}

For completeness, we list also the explicit expressions for the entropy. It
includes contribution $S_{0}$ from the horizon \cite{frolov}, \cite{gibper}
(2D analog of Bekenstein-Hawking entropy) and quantum corrections $S_{q}$ 
\cite{qent}, \cite{qent2}$.$ The constant in $S_{q}$ is chosen to ensure $%
S_{q}\rightarrow 0$ when the boundary value $r_{B}\rightarrow r_{+}$ (no
room for radiation). 
\[
S=S_{0}+S_{q}\text{, }S_{0}=2F(\phi _{+})\text{.} 
\]

\begin{equation}
S_{q}=2\kappa (\psi _{B}-\psi _{+})=-2\kappa \int_{\phi _{+}}^{\phi _{B}}%
\frac{[f_{\phi }^{(0)\prime }-f_{+}^{(0)\prime }]}{f}
\end{equation}
\begin{equation}
S_{q}=\kappa s_{q}\text{, }s_{q}=2(\phi _{+}-\phi _{B})(1-x)+2(1+x)\ln \frac{%
1-x}{1-y}\text{.}
\end{equation}

In the limit $r_{B}\gg r_{+}$ we have 
\begin{equation}
s_{q}=2(\phi _{+}-\phi _{B})(1-x)+2(1+x)\ln (1-x)\text{.}  \label{sq}
\end{equation}
The first term looks like the contribution of hot thermal gas in a flat
space with the temperature $T_{H}$ (with quantum correction omitted since $%
s_{q}$ is itself multiplied by $\kappa $)$.$ The second term may be
interpreted as the contribution from vacuum polarization and diverges in the
extremal limit $x\rightarrow 0$.

\subsection{First law}

Having explicit expressions for the quantum-corrected mass, entropy,
temperature and potential one can check by direct calculations that the
general first law holds with quantum corrections taken into account:

\[
\delta M_{tot}=T_{H}\delta S_{tot}-\varphi _{+}\delta Q\text{.} 
\]

\section{Quantum-corrected extreme configuration}

Up to now, we were dealing with non-extremal black holes. In this section we
start from the very beginning from the configuration which belongs to
another topological class - extremal black holes ($T_{H}=0$), and find
quantum corrections to the mass. Afterwards, we compare the result with what
is obtained by the limiting transition from the NEBH to the EBH and show
that the results coincide.

As is shown in \cite{ecq}, the metric describing a static black hole
(non-extremal or extremal) can be represented in the form which includes
only derivatives with respect to the dilaton (with the coordinate $x$
removed):

\begin{equation}
f=\chi \exp (-\gamma )\text{, }\chi ^{\prime }=U\tilde{w}\exp (-\gamma )%
\text{, }  \label{ff}
\end{equation}

\begin{equation}
\alpha =\frac{V-\frac{\kappa y^{2}}{2}}{w-\kappa y}\text{, }  \label{al}
\end{equation}
\[
\chi =\chi _{0}-\kappa \chi _{1}\text{, }\chi _{0}=\int_{\phi _{+}}^{\phi
}d\phi U\exp (-\gamma )w\text{,} 
\]
\[
\chi _{1}=\int_{\phi _{+}}^{\phi }d\phi U\exp (-\gamma )y 
\]
\[
\gamma =\int d\phi \alpha \text{, }\alpha \equiv \frac{\tilde{V}}{\tilde{w}}%
\text{, }\tilde{w}=w-\kappa y\text{, }y=\psi ^{\prime }\text{, }w=F^{\prime
} 
\]

From now on we put for simplicity $b=c=0$. Then $w=-\exp (-\phi )$, 
\begin{equation}
\alpha =-\frac{1-\frac{\kappa y^{2}}{2}\exp (\phi )}{1+\kappa y\exp (\phi )}%
\text{.}  \label{aa}
\end{equation}

In the pure classical case $\kappa =0$ and eqs. (\ref{ff})-(\ref{aa})
actually give us exact solutions in agreement with the known fact that
classically any gravitation-dilaton 2D system is integrable. However, for $%
\kappa \neq 0$ we have non-linear integral-differential equations since $f$
enters the right hand side via the quantity $y$. We may try to solve them
approximately, putting in the right hand side $y=y^{(0)}$ in terms with the
small parameter $\kappa $, where $y^{(0)}$ corresponds to the classical
extremal solution ($\kappa =0$) for a given value of $\phi _{+}$: $%
f^{(0)}=[1-\exp (\phi -\phi _{+})]^{2}$, so 
\[
y^{(0)}=\frac{2u}{1-u}\text{, }u\equiv \exp (\phi -\phi _{+}) 
\]

\[
U=\exp (-\phi )[1-q^{2}\exp (2\phi )]. 
\]

The condition of extremality (\ref{cond}) gives us 
\begin{equation}
\exp (\phi _{+})=q+\kappa \text{,}  \label{condq}
\end{equation}
whence 
\[
U=\exp (-\phi )(1-u^{2})+2\kappa u\text{,} 
\]

\[
\alpha =\frac{V-\frac{\kappa y^{2}}{2}}{w-\kappa y}\text{,} 
\]
\[
\chi =\chi _{0}-\kappa \chi _{1}\text{, }\chi _{0}=\int_{\phi _{+}}^{\phi
}d\phi U\exp (-\gamma )w\text{,} 
\]
\[
\chi _{1}=\int_{\phi _{+}}^{\phi }d\phi U\exp (-\gamma )y\text{,} 
\]
\[
\gamma =\int d\phi \alpha \text{, }\alpha \equiv \frac{\tilde{V}}{\tilde{w}}%
\text{, }\tilde{w}=w-\kappa y\text{, }y=\psi ^{\prime }\text{, }w=F^{\prime }%
\text{.} 
\]

We have in the approximation under consideration: 
\[
\gamma =-\int duT\text{, }T=\frac{(1-u)^{2}-2\tilde{\kappa}u^{3}}{%
u(1-u)(1-u+2\tilde{\kappa}u^{2})}\text{, }\tilde{\kappa}\equiv \kappa \exp
(\phi _{+})\text{.} 
\]
After integration, we obtain neglecting terms which contains $\kappa ^{2}$:

\[
\exp (-\gamma )=\frac{(1-u)}{u_{1}-u}u\exp (\phi _{+})[1-2\tilde{\kappa}\ln
(1-u)]\text{,} 
\]
where $u_{1}=1+2\tilde{\kappa}$, the constant of integration is chosen to
ensure $\gamma =-\phi $ in the classical limit $\kappa =0$.

Then after simple but somewhat lengthy calculations we obtain 
\begin{equation}
f=(1-u)^{2}+\tilde{\kappa}e(u)  \label{fe}
\end{equation}
\begin{equation}
e(u)=-4(1-u)^{2}\ln (1-u)-u(1-u)^{2}  \label{ex}
\end{equation}

Now we can check the metric of the quantum-corrected EBH (\ref{fe}), (\ref
{ex}) with the limiting form of the NEBH metric (\ref{f})-(\ref{hi}). In
doing so, one should be careful since in (\ref{f})-(\ref{hi}) quantum
corrections stem not only from the terms with $\chi $, but also from the
term $f^{(0)}$. This happens since it contains the parameter $r_{-}=\frac{%
q^{2}}{r_{+}}$ in which $\kappa $ appears due to the extremality condition (%
\ref{condq}) relating $q$ and $r_{+}$. With this in mind, one can check that
the results of implementing the extremality conditions to (\ref{f})-(\ref{hi}%
) do coincide with eqs. (\ref{fe})-(\ref{ex}). Thus, the metric of the
non-extremal configuration tends to that of the extremal one in the limit
under discussion.

For small $u$

\[
f=1-B\exp (\phi -\phi _{+})\equiv 1-Bu\text{, }B=2-3\tilde{\kappa}\text{.} 
\]

As is shown in \cite{ecq}, it follows from field equations that

\begin{equation}
\frac{\partial \phi }{\partial z}=z_{0}\tilde{w}^{-1}e^{\gamma }\text{, }%
z_{0}=const\text{.}  \label{qh}
\end{equation}
In the main approximation 
\[
\frac{d\phi }{dz}=-\frac{B}{1+\kappa y\exp (\phi )}\frac{(u_{1}-u)}{1-u}[1+2%
\tilde{\kappa}\ln (1-u)] 
\]
At infinity the quantum corrections vanish and we have the linear dilaton
vacuum:$\frac{\partial \phi }{\partial z}=-1$, whence 
\[
B=\frac{1}{u_{1}}=1-2\tilde{\kappa} 
\]
and 
\[
\frac{d\phi }{dz}=-1+\tilde{\kappa}O(u^{2})\text{, }u\ll 1\text{.} 
\]

Applying now eq. (\ref{e}), one can calculate the energy at infinity,
knowing expansions for $f(\phi )$ and $\frac{\partial \phi }{\partial z}$
and subtracting the energy of the corresponding quasiflat configuration.
Calculations go along the same line as in the non-extremal case and give us 
\[
m=\exp (-\phi _{+})-\frac{3}{2}\kappa . 
\]
With the same accuracy 
\[
m=q-\frac{\kappa }{2}\text{, }m<q\text{.} 
\]

Comparing the expression for the energy of the near extremal BH, taking into
account the extremality condition and neglecting terms $\kappa ^{2}$, we see
that 
\[
m=\frac{\exp (-\phi _{+})+q^{2}\exp (\phi _{+})}{2}-\frac{\kappa }{2}%
\rightarrow \exp (-\phi _{+})-\frac{3}{2}\kappa \text{.} 
\]

Thus, the mass of the quantum-corrected non-extremal configuration tends to
that of the pure extreme one when the relationship between $q$ and $\phi
_{+} $ approaches the extremality condition (\ref{cond}). It is worth noting
that the account for backreaction of massive fields in general relativity
also gives the result that for the quantum-corrected extremal charge black
hole $m<q$ \cite{jurek}.

Thus, the relationship between mass and charge of the extremal configuration
is situated in the space of parameters in the region where the very
existence of black holes is forbidden classically.

\section{Extreme limit of quantum-corrected non-extreme configuration}

The formulas for the quantum-corrected metric (\ref{fe}), (\ref{ex}) fail in
the immediate vicinity of the horizon. Indeed, in this vicinity $%
y\rightarrow \infty $, while the coefficients $V(\phi _{+})$, $w(\phi _{+})$
are finite. Therefore, near the horizon the terms of the type $\kappa y$ or $%
\kappa y^{2}$ cannot be considered in equation (\ref{al})  as small
corrections and, moreover, they dominate the corresponding equations. The
asymptotic form of the metric of a generic quantum-corrected extremal black
hole was found in \cite{ecq} that generalized the previous result of \cite
{triv} for the particular model. It turned out that 
\begin{equation}
f\sim (\phi -\phi _{+})^{\varepsilon }\text{,}  \label{ef}
\end{equation}
where for small $\kappa $ the quantity $\varepsilon =2+\kappa \nu $, the
coefficient $\nu $ (whose exact form is irrelevant for us now) depends on
the behavior of $V$ and $F$ near the horizon. It is essential that the
dependence of the metric on dilaton and that of dilaton on the Schwarzschild
coordinate $z$ is non-analytical in this region. On the other hand, it is
obvious that for the non-extremal configuration, even if it is close to the
extremal one as nearly as one likes, the metric is analytical there. How it
may happen and what is going with the non-extremal metric in the extremal
limit?

Let us trace the transition under discussion in more detail. For any
non-extremal configuration we may exploit the power expansion near the
horizon: 
\begin{equation}
f=a_{1}(z-z_{+})+a_{2}(z-z_{+})^{2}+a_{3}(z-z_{+})^{3}+...\text{.}
\end{equation}
\begin{equation}
\phi =\phi _{+}+b_{1}(z-z_{+})+b_{2}(z-z_{+})^{2}+b_{3}(z-z_{+})^{3}+...
\label{d1}
\end{equation}
It is convenient to use eqs. (\ref{3}) and the dilaton equation which can be
obtained by variation with respect to $\phi $: 
\begin{equation}
RF^{^{\prime }}+U^{\prime }=2V\square \phi +V^{\prime }(\nabla \phi )^{2}.
\end{equation}
Substituting into them (\ref{f1}) and (\ref{d1}), taking into account that $%
R=-f^{\prime \prime }$ and equating coefficients at equal powers of $z-z_{+}$%
, we obtain the relations between coefficients. In the limit $%
a_{1}\rightarrow 0$ (extreme limit) they simplify to give 
\begin{equation}
U^{\prime }(\phi _{+})=2a_{2}F^{\prime }\left( \phi _{+}\right) \text{,}
\end{equation}
\begin{equation}
b_{1}[U^{\prime }(\phi _{+})-2a_{2}F^{\prime }\left( \phi _{+}\right)
]=6\kappa a_{3}\text{,}
\end{equation}
whence we obtain 
\begin{equation}
\kappa a_{3}=0\text{.}
\end{equation}

For the classical system, when $\kappa =0$ exactly, this turns into identity
and does not impose any restriction on $a_{3}$. Continuing the procedure
order by order, one may restore the whole expansion which can be also
obtained directly from the exact classical expression for the extremal black
hole $f^{(0)}=[1-\exp (\phi -\phi _{+})]^{2}$. However, for the
quantum-corrected case the situation is qualitatively different. Whatever
small $\kappa $ be, it entails immediately $a_{3}=0$. Continuing the
procedure, one can easily obtain that all $a_{n}=0$ with $n\geq 3$. In a
similar way, all $b_{n}=0$.

Let me remind that, apart from the main branch of solution, $\phi (z)$, for
the gravitation-dilaton system there also exists the special one $f_{c}$
with $\phi =\phi _{0}=const$ (2D analogs of Bertotti-Robinson spacetime).
For exactly solvable models the values $\phi _{0}=\phi _{s}$ correspond to
the singularity of the main branch, \cite{exact}, \cite{sol96}, while in the
generic case $\phi _{s}$, \cite{ecq}, \cite{deg} coincides with the value
typical of the extremal horizon. This second branch manifests itself just now%
$.$

Thus, in the vicinity of the horizon ($z\rightarrow 0$) in the extremal
limit ($T_{H}\rightarrow 0$) 
\[
f_{n}(z,\kappa ,T_{H})\rightarrow f_{c}(z\text{, }\kappa ) 
\]
but not to the quantum-corrected extremal solution $f_{e}$.

Meanwhile, for the pure classical system 
\begin{equation}
f_{n}(z,0\text{, }T_{H})\rightarrow f_{e}(z,0)  \label{fzf}
\end{equation}
smoothly for all $z$.

Beyond the immediate vicinity of the horizon the situation is qualitatively
different: as we saw, (\ref{fzf}) holds true.

The size of this vicinity is governed by the parameter $k=\kappa \left| \ln
(\phi -\phi _{+})\right| $. If $k\ll 1$, the additional factor $(\phi -\phi
_{+})^{\kappa \nu }$in the metric function due to quantum corrections is
close to $1$ and the relationship between the geometry of NEBH and EBH is
similar to that for the pure classical case. However, if $k\gtrsim 1$, the
quantum corrections changes the picture significantly and should be taken
into account from the very beginning that just leads to the expression like (%
\ref{ef}).

\section{Spherically-symmetrical reduction and quantum corrections}

In this section we consider another physically relevant 2D model of dilaton
gravity which classically appears within the framework of SSG. As in the
previous sections, it is assumed that the quantum backreaction is due to
minimal fields, so it is described by the Polyakov-Liouville action (\ref
{ipl}). Quantum corrections for such a system were studied in \cite{sol3}
for NEBH. Therefore, we restrict ourselves by the case of EBH only. Now the
action looks like (\ref{ac}) with coefficients

\[
F=r^{2}\text{, }w=2\phi \text{, }V=2\text{, }U=2(1-\frac{Q^{2}}{r^{2}})\text{%
, }r=\phi \text{,} 
\]
and the common factor $\frac{1}{4}$ instead of $\frac{1}{2\pi }$ (see \cite
{sol3} for details). As calculations run almost along the same lines as for
the model discussed in the text, above, I list basic formulas and results
only.

For the classical extremal black hole $f_{ext}^{(0)}=\left( 1-\frac{r_{+}}{r}%
\right) ^{2}$, so $y^{(0)}=\frac{2r_{+}}{r^{2}\left( 1-\frac{r_{+}}{r}%
\right) }$. The extremality condition (\ref{cond}) gives now 
\begin{equation}
Q=r_{+}-\frac{\kappa }{2r_{+}}\text{.}  \label{cond2}
\end{equation}
The quantum-corrected metric function has the form (\ref{fe}), where now 
\[
e(u)=-4+5u-u^{2}-\frac{4(1-u)^{2}}{u}\ln (1-u)\text{.} 
\]
It follows from (\ref{qh}) that at large distances $\frac{\partial \phi }{%
\partial r}=1+\kappa O(\frac{1}{r^{3}})$. Then, calculating the energy at
infinity we obtain after some algebra that 
\[
M_{tot}=E-E_{0}=r_{+}+\frac{\kappa }{2r_{+}}\text{,} 
\]
where $E_{0}$ is the contribution of the flat reference background. Taking
into account the extremality condition (\ref{cond2}) we obtain 
\[
M_{tot}=Q+\frac{\kappa }{r_{+}}>Q\text{.} 
\]

Thus, in contrast to the charged CGSH black hole, now the extremal
quantum-corrected black hole lies inside the classically allowed region of
parameters.

\section{Summary}

We have found quantum corrections caused by minimal fields to the geometry
and thermodynamics of non-extremal charged black holes in the Hartle-Hawking
state for string-inspired dilaton theories of gravity that include in the
uncharged case CGHS black holes. In the limit of the zero charge the quantum
corrections for the non-extremal black hole agree with our previous
calculations \cite{2dwall}. The validity of the first general law is
demonstrated with quantum corrections taken into account.

We have shown that the mass splits in two pieces. The first one ($M_{BH}$)
is the mass of the black hole itself that contains the horizon dilaton value 
$\phi _{+}$ but does not depend on the dilaton field in the point of
observation in an asymptotically flat region $\phi _{B}.$ The second one ($%
M_{th}$) is proportional to the difference $\phi _{+}-\phi _{B}$ and
coincides with that of a thermal gas at the temperature $T_{H}$ in a box of
corresponding size in the flat space. In the extremal limit $%
M_{th}\rightarrow 0$, while $M_{BH}$ remains finite. Meanwhile, the quantum
correction to the entropy (\ref{sq}) diverges in the extremal limit that
agrees with previous observations \cite{gab}, \cite{bur} where it was
attributed to the failure of the one-loop approximation. However, in our
view, this does not mean necessarily that such failure happens. Rather, it
looks like the consequence of the changes in geometry: the bulk contribution
from $S_{q}$ becomes divergent because of the fact that the proper distance
between the horizon and any other point tends to infinity in the extremal
limit. We also obtained, in contrast to \cite{gab}, the finite shift in the
horizon value of the dilaton for EBH that is described for a given charge by
very simple formulas (\ref{condq}), (\ref{cond2}). The reason of discrepancy
is not quite clear. We only suppose that it is somehow connected with the
different definitions of the unperturbed state in the situation of
competition between corrections of two kinds connected with quantum effects
and small deviation from the extremality.

The results of the present paper along with the previous one \cite{ecq}
confirm the existence of semiclassical 2D EBH black holes for massless
minimal fields. We traced the limiting transition from the quantum-corrected
non-extremal black hole to the extremal one and showed that far from the
horizon the geometry and mass of the first type configuration tend smoothly
to those of the second one. However, in the vicinity of the horizon the
situation changes: in the limit under discussion the solution tends to that
with the constant dilaton that represents the 2D analogue of the
Bertotti-Robinson spacetime. Meanwhile, for the pure classical black hole
the limiting transition is uniform in the sense that non-extremal
configurations go smoothly to the extremal one everywhere.. Thus, there is a
crucial difference in this respect between classical and quantum-corrected
black hole configurations. This phenomenon, along with the non-analytical
behavior of the extremal metric on the Schwarzschild coordinate \cite{triv}
show that quantum backreaction becomes crucial for properties of extremal
black holes. Meanwhile, it does not affect their very existence but only
changes slightly the condition of extremality.

Calculation of the mass of the quantum-corrected EBH showed that, depending
on the model, $M_{tot}<Q$ or $M_{tot}>Q$. It is worth mentioning that $%
M_{tot}<Q$ for backreaction of massive fields on 4D quantum-corrected
extremal Reissner-Nordstr\"{o}m black hole \cite{jurek}.

Thus, in the simplified 2D context we manage to trace how quantum
backreaction change properties of near-extremal and extremal black holes.

\section{Acknowledgment}

I thank for hospitality Erwin Schr\"{o}dinger International Institute for
Mathematical Physics where this work started at the Wokshop on 2D gravity. I
am also grateful to Daniel Grumiller for discussions and stimulating
interest.




%
%

%
%

\end{document}